\documentclass{nature}

\usepackage[top=0.8in, bottom=1in, left=0.8in, right=0.8in]{geometry}

\usepackage{graphicx}
\usepackage{amsmath}
\usepackage{amssymb}

\newcommand{\msun}{M$_{\odot}$}

\title{A massive compact quiescent galaxy at z\,=\,2 with a complete Einstein ring in JWST imaging
}

\author{\large Pieter van Dokkum$^{1}$,
Gabriel Brammer$^{2,3}$,
Bingjie Wang$^{4}$,
Joel Leja$^{4,5}$,
Charlie Conroy$^{6}$
\vspace{8pt}}

\begin{document}

\maketitle

\begin{affiliations}
\small
 \item  Department of Astronomy, Yale University, New Haven, CT 06511, USA
 \item  Cosmic Dawn Center (DAWN), Denmark 
 \item Niels Bohr Institute, University of Copenhagen, Jagtvej 128, DK-2200 Copenhagen N, Denmark
 \item  Department of Astronomy \& Astrophysics, The Pennsylvania State University, University Park, PA 16802, USA
 \item Institute for Computational \& Data Sciences, The Pennsylvania State University, University Park, PA 16802, USA
\item Harvard-Smithsonian Center for Astrophysics, 60 Garden Street, Cambridge, MA, USA
 
\end{affiliations}


\begin{abstract}
One of the surprising results from HST was the discovery that many 
of the most massive galaxies at $z\sim 2$ are very compact, having
half-light radii of only $1-2$\,kpc.
The interpretation is that massive galaxies formed
inside-out, with their cores largely in place by $z\sim 2$
and approximately half of their present-day mass added later
through minor mergers.
Here we present a compact, massive, quiescent galaxy at
$z_{\rm phot}=1.94^{+0.13}_{-0.17}$ with a complete Einstein ring.
The ring was found in the JWST COSMOS-Web survey and is
produced by a background galaxy at $z_{\rm phot}=2.98^{+0.42}_{-0.47}$.
Its $1.54\arcsec$ diameter provides a
direct measurement of the mass of the ``pristine''
core of a massive galaxy, observed before the
mixing and dilution of its stellar population
during the 10\,Gyr of galaxy evolution between
$z=2$ and $z=0$.
We find a mass of $M_{\rm lens}=6.5^{+3.7}_{-1.5}
\times 10^{11}$\,\msun\ within
a radius of 6.6\,kpc. The stellar mass within the same radius is
$M_{\rm stars}= 1.1^{+0.2}_{-0.3} \times 10^{11}$\,\msun\ for a Chabrier
initial mass function (IMF),
and the fiducial dark matter
mass is $M_{\rm dm} = 2.6^{+1.6}_{-0.7}
\times 10^{11}$\,\msun.
Additional mass
is needed to explain the lensing results,
either in the form of a 
higher-than-expected dark matter density or a bottom-heavy IMF.

\end{abstract}

\begin{figure*}[htbp]
  \begin{center}
  \includegraphics[width=0.9\linewidth]{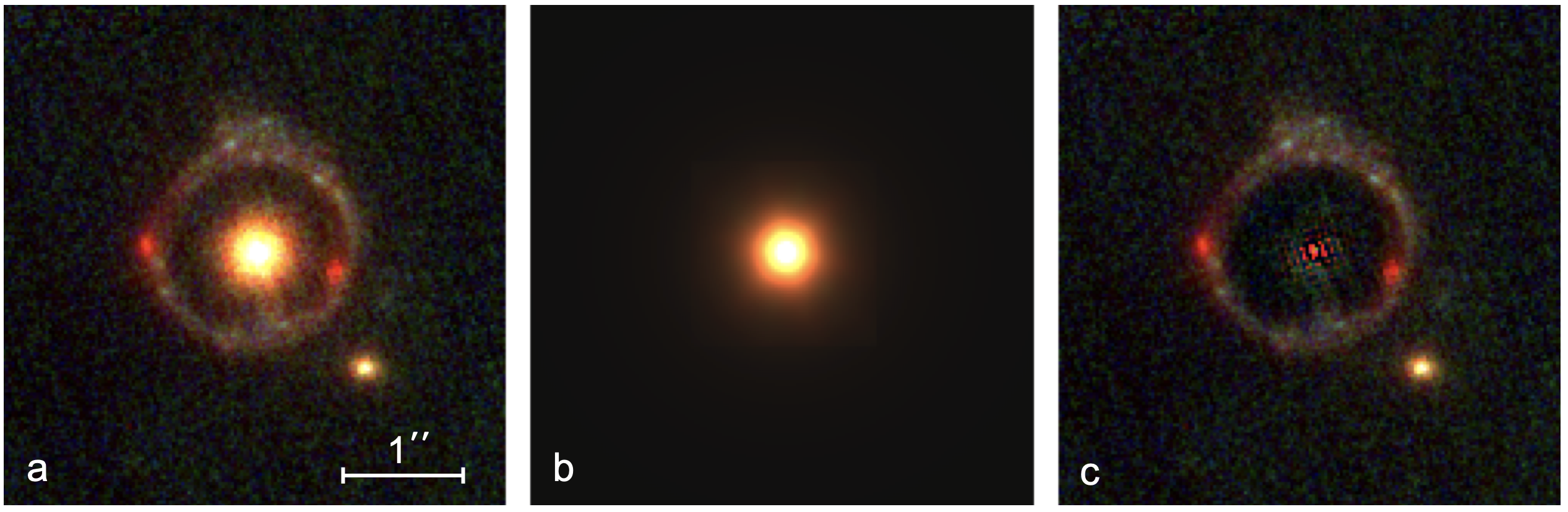}
  \end{center}
    \vspace{-0.3truecm}  
    \caption{\small \textbf{A complete Einstein ring identified in JWST images.} 
{\em a)} Color image of JWST-ER1, created from the NIRCam F115W, F150W,
and F277W data. {\em b)} Model of the galaxy, with an effective radius
of $r_{\rm e}=1.9$\,kpc. {\em c)} Residual of the fit. Each panel spans
$4.1''\times 4.1''$. The coordinates of the lens are ${\rm RA}=10^{\rm h}00^{\rm m}24.11^{\rm s}$, ${\rm Dec}=01^{\circ}53'34.9''$ (J2000).
   }
   \label{mainfig.fig}
    \vspace{-12pt}
\end{figure*}

The galaxy and its ring were identified in JWST
NIRCam observations in the context of the
COSMOS-Web project,\cite{cosmosweb}
a public wide-area survey using the F115W, F150W, F277W, and F444W filters.  A visual inspection of a mosaic generated from
the F115W, F277W, and F444W data available as of 15 Apr 2023, covering a total area of 0.35\,deg$^2$, readily revealed the object (see Methods).
The NIRCam images containing the galaxy
were resampled\cite{fruchter:02} to a common $0.025''$\,pix$^{-1}$ grid for analysis.

The object, dubbed JWST-ER1, is shown in Fig.\
\ref{mainfig.fig}a. It consists of a compact early-type galaxy
(JWST-ER1g) and a complete Einstein ring (JWST-ER1r) with two conspicuous
red concentrations. The lensed galaxy likely has a red center and a
blue disk, with parts of the disk producing the ring. The diameter
of the center of the ring is $1.54''\pm 0.02''$.
JWST-ER1 joins a large number of known Einstein rings,\cite{lehar:93,bolton:06} although most are not complete. Like other strong lensing configurations
Einstein rings can be used to reconstruct high resolution
images of lensed background galaxies, using ray tracing
techniques.\cite{autolens}  However, the unique value
of Einstein rings is what they tell us about the lenses themselves:
given the redshifts of the lens and source,
they provide a model-independent measurement of the enclosed mass
within the radius of the ring.\cite{kochanek:01}

\begin{figure*}[hbtp]
  \begin{center}
  \includegraphics[width=0.95\linewidth]{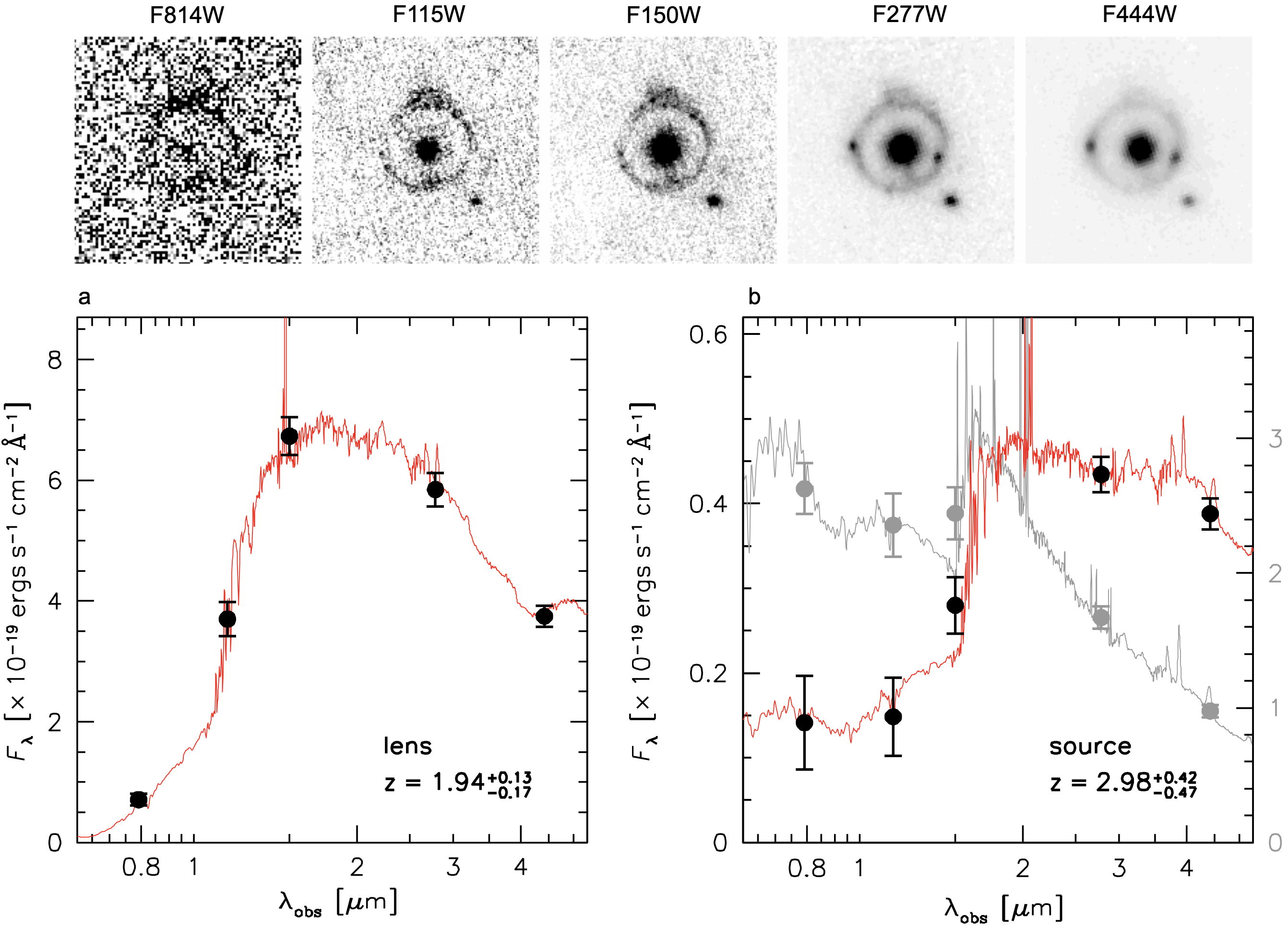}
  \end{center}
    \vspace{-0.3truecm}  
    \caption{\small \textbf{Photometry of the lens and source.} 
Top panels: images in the HST/ACS F814W band and the JWST/NIRCam
F115W, F150W, F277W, and F444W bands. They are shown at a common
$F_{\lambda}$ scale. {\em a)} SED of the lens galaxy, determined
from forced {\tt galfit} fits. The galaxy is well-fit by a
quiescent stellar population at $z=1.94^{+0.13}_{-0.17}$
and a total stellar mass of $1.1^{+0.3}_{-0.4}\times 10^{11}$\,\msun\ (for a Chabrier IMF). {\em b)}
SED of the lens galaxy, with the summed flux of the two red knots
shown in black and the blue ring in grey. The red knots provide
a reasonably well-constrained redshift of
$z_{\rm phot} = 2.97^{+0.44}_{-0.37}$. Data are presented as
measurements $\pm$\,sd.
   }
   \label{sed.fig}
    \vspace{-12pt}
\end{figure*}

We obtained five-band photometry of the lens by fitting it with a
S\`ersic model,\cite{sersic}
masking the ring and keeping the structural parameters fixed in all bands.
The effective radius of the galaxy $r_{\rm e}=0.22'' \pm 0.02''$ and
its Sersic index $n=5.0\pm 0.6$. The total magnitudes of the
galaxy are given in Table 1 and the spectral energy distribution
(SED) is shown in Fig.\ \ref{sed.fig}a.
There is a pronounced break between the F814W and F115W bands,
leading to a well-constrained photometric redshift of
$z=1.94^{+0.13}_{-0.17}$ for the lens (see Methods). The photometric
redshift exceeds the spectroscopic redshift of
the most distant known lens, a $z=1.525$ star
forming galaxy with a complex morphology.\cite{canameras:17}
The source redshift is less well constrained. We split the
source into two photometric masks, one containing the blue ring and
one covering both of the red knots. The blue ring shows no strong
features, and has a redshift of
$z_{\rm phot}=2.89^{+0.27}_{-0.98}$. The SED of the red knots has a clear break
between F150W and F277W, and a better-constrained
redshift of $z_{\rm phot}=2.98^{+0.42}_{-0.47}$
(see Fig.\ \ref{sed.fig}b).

\begin{figure*}[htbp]
  \begin{center}
  \includegraphics[width=0.85\linewidth]{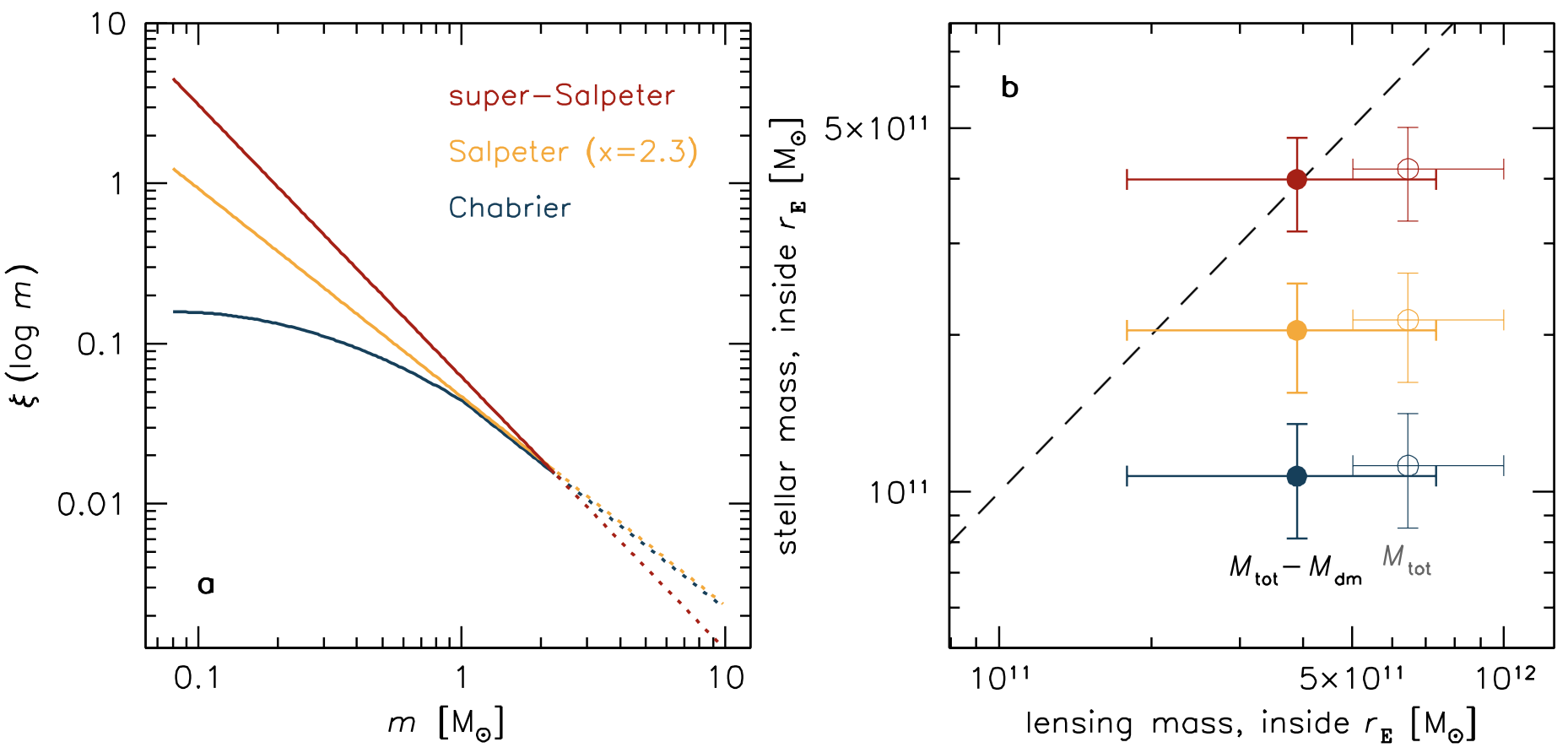}
  \end{center}
    \vspace{-0.3truecm}  
    \caption{\small \textbf{Comparison of stellar mass to lensing mass for
    different IMFs.} 
{\em a)} The three IMFs that are considered in this study: a Chabrier
IMF,\cite{chabrier:03} a Salpeter IMF,\cite{salpeter:55} and
an IMF that is steeper than Salpeter with a slope of $-2.7$. Broken lines are for stars above the
turn-off mass.
{\em b)} Comparison of the stellar mass to the lensing mass (open
symbols) and the lensing mass minus the fiducial amount of
dark matter (solid symbols), for the three IMFs. The dashed line indicates a one-to-one relation. Data points are measurements $\pm$\,sd.
   }
   \label{constraints.fig}
    \vspace{-12pt}
\end{figure*}

The lensing galaxy appears to be a textbook example of the class of 
massive quiescent galaxies at $z\sim 2$. Its rest-frame
colors, $U-V\approx 2.10$ and $V-J\approx 1.3$, place it
comfortably in the quiescent region of the $z\sim 2$
UVJ diagram.\cite{whitaker:11} The best fit stellar population
parameters from the Prospector\cite{johnson:21} fit imply an age of $1.9^{+0.3}_{-0.6}$\,Gyr
and a low star formation rate of $4^{+19}_{-3}$\,\msun{}yr$^{-1}$.
The Prospector total stellar mass of JWST-ER1g
is $1.3^{+0.3}_{-0.4} \times 10^{11}$\,\msun\ for a Chabrier\cite{chabrier:03}
IMF,
and its observed effective radius corresponds to $r_{\rm e}=1.9\pm
0.2$\,kpc. This makes the galaxy quite compact, just like other
quiescent galaxies at these redshifts,\cite{daddi:05,trujillo:06,dokkumnic:08,barro:13} and it falls on
of the canonical size-mass relation of quiescent
galaxies.\cite{wel:14} The
galaxy is almost perfectly round and there are no obvious 
star forming regions, tidal tails, or other irregularities in
the residuals from the {\tt galfit} fit.

We now turn to the mass of JWST-ER1g as inferred from
the radius of the Einstein ring.
The photometric redshifts of the lens and source, combined with the
radius of the Einstein ring, give a total mass of $M_{\rm lens}
=6.5^{+3.7}_{-1.5} \times 10^{11}$\,\msun\ within
$r=6.6$\,kpc (see Methods). The stellar mass within the
Einstein radius is $0.79\times$ the total mass as determined
by {\tt galfit} and {\tt Prospector}, that is, $M_{\rm stars}=
(1.1^{+0.2}_{-0.3}\times 10^{11}$\,\msun\ for a Chabrier IMF.
There is a large difference between the lens mass
and the Chabrier stellar mass of JWST-ER1g, with the lens
mass a factor of $5.9^{+4.1}_{-1.6}$ higher than the stellar mass.
This is the central result of our study (besides the report of the discovery of JWST-ER1), and in the following
we discuss several possible contributors to the lensing mass.

It is unlikely that a significant
fraction of the lensing mass is in the form
of gas. Observations of lensed quiescent galaxies,\cite{whitaker:21nat}
as well as simulations,\cite{johansson:12,whitaker:21} have consistently found
low gas masses ($<10^{10}$\,\msun)
for massive quiescent galaxies at these redshifts.
Furthermore,
a total gas mass of $3\times 10^{11}$\,\msun\ within 6.6\,kpc corresponds to
such a high  projected gas density that a high star formation rate
is inevitable. The average projected
surface density would be $\approx 2200$\,\msun\,pc$^{-2}$, and
according to the Kennicutt-Schmidt relation\cite{ks:98} the corresponding
star formation rate surface density is $\Sigma
\approx 14$\,\msun\,yr$^{-1}$\,kpc$^{-2}$. The total star formation
rate within the ring would be ${\rm SFR}\approx 2000$\,\msun\,yr$^{-1}$,
three orders of magnitude higher than derived from the {\tt Prospector}
fits and $30\times$ higher than an upper limit derived from
Spitzer/MIPS 24\,$\mu$m data (see Methods).
This is a rough estimate, with the
actual SFR depending on the distribution and temperature
of the gas, but the point is that JWST-ER1g would not be quiescent but
a strong starburst galaxy.

There is of course dark matter
within the Einstein ring, and with standard assumptions this
explains about half of the difference between the lensing mass
and the stellar mass. Assuming
an NFW profile\cite{navarro:97} and the stellar mass -- halo mass relation\cite{behroozi:13b}
for $z=2$, the dark matter mass
within the Einstein radius is $M_{\rm dm}=2.6^{+1.6}_{-0.7}\times 10^{11}$\,\msun\
(see Methods). As shown in Fig.\ \ref{constraints.fig} this leaves $2.8^{+3.4}_{-2.1}\times 10^{11}$\,\msun\ unaccounted for.
An explanation for this mild discrepancy
is that the dark
matter density within the Einstein radius
is a factor of $\sim 2$ higher than expected
from scaling relations.
The ``extra'' dark matter can come in two forms. First, the total halo mass could be higher than what is indicated by the canonical stellar mass -- halo
mass relation. 
A second option is that
baryonic processes have led to a dark matter profile that deviates from the NFW
form. The final profile can be steeper or shallower in the central regions,
depending on the balance between 
cooling and feedback.\cite{duffy:10,schaller:15,tollet:16}

Looking closer, both options are somewhat unlikely in the
specific case of JWST-ER1g. As detailed
in the Methods section the total halo mass would have to be
very high, close to $M_{\rm halo} \sim 10^{14}$\,\msun, and
only a few halos of that
mass are expected to exist in the surveyed volume.
Turning to baryonic processes, they
tend to alter the dark matter profile on the spatial scales where the baryons are: specifically, significantly steeper profiles are expected in regions where the stellar mass dominates,\cite{schaller:15} that is, at radii 
$\lesssim r_{\rm e}$. The
dark matter mass within 1.9\,kpc is only $3\times 10^{10}$\,\msun\ for a $10^{13}$\,\msun\ halo with an
NFW profile, and even if this were enhanced by a factor of 2--3 it would not be
enough to account for the missing mass within the Einstein radius.

An intriguing alternative is that the missing mass is in the form of low
mass stars, and that the stellar IMF needs to be adjusted:
stars with masses $M\sim 0.5$\,\msun\
and below dominate the total mass but contribute less than 5\,\% to
the light.\cite{dokkum:21chromo} 
Rather than simply scaling the mass,
we refit the photometry in Prospector with
two bottom-heavy IMFs: the Salpeter
form,\cite{salpeter:55} with a slope of $-2.3$ and no turnover, and
a ``super-Salpeter'' IMF with a slope of $-2.7$.
These IMFs are illustrated in Fig.\ \ref{constraints.fig}a.
We note that these parameterizations are not unique, as the low mass slope is degenerate with the low mass cut-off. Furthermore, top-heavy IMFs can lead to high $M/L$ ratios too if the mass is dominated by stellar remnants, although even for very flat IMFs this only occurs at ages $>3\times 10^9$\,Gyr.\cite{maraston:98}
With these caveats in mind, we find that the stellar mass within the Einstein radius is $2.0^{+0.5}_{-0.5} \times 10^{11}$\,\msun\ for a
Salpeter IMF and $4.0^{+0.6}_{-0.8} \times 10^{11}$\,\msun\ for the super-Salpeter IMF. 
As shown in Fig.\ \ref{constraints.fig}b a model that combines
a super-Salpeter IMF with a standard dark matter halo matches
the lensing mass exactly, with a Salpeter IMF also providing a good fit.


The likely descendants of compact quiescent galaxies at $z\sim 2$ are 
massive early-type
galaxies,\cite{bezanson:09,naab:09inside,dokkumgrow:10,sande:13} and the central regions of these galaxies
may indeed have IMFs that are more bottom-heavy than the
Chabrier IMF. The evidence largely comes from gravity-sensitive absorption
lines,\cite{conroy:imf12} kinematics,\cite{cappellari:12} and gravitational
lensing.\cite{treu:10} Outside of the central regions there appears
to be a gradual
transition to a Chabrier IMF,\cite{martinnavarro:15,labarbera:16,dokkum:17imf,smith:20}
as expected if a significant fraction
of the mass in the outskirts was accreted through minor mergers.
Quantitatively, the excess stellar mass compared to a Chabrier IMF reaches
a factor of $\sim 3$ in the centers of massive galaxies,
with a powerlaw slope of $-2.7$ found for the galaxy NGC\,1407 from a
detailed non-parametric analysis.\cite{conroy:17} Super-Salpeter
slopes of $-2.7$ have also been proposed on theoretical grounds.\cite{chabrier:14}
We infer that a steep IMF
for JWST-ER1  would be consistent with
estimates in the central regions of early-type galaxies, particularly
when mixing and dilution due to mergers and projection effects are taken
into account.\cite{sonnenfeld:17}

While this consistency is encouraging,
IMF measurements are difficult and often
indirect, and the question of IMF variation in the central regions of elliptical
galaxies is still debated.\cite{smith:15,smith:20} 
Furthermore, and of direct relevance to JWST-ER1g,
bottom-heavy IMFs are in some tension with comparisons
of dynamical masses to stellar masses of $z\sim 2$
galaxies, which tend to prefer bottom-light IMFs such
as the Chabrier form.\cite{belli:14,esdaile:21}
On the other hand,
our results are qualitatively consistent with the
most similar system to JWST-ER1, a
$z=1.525$ lens and partial Einstein ring
that is best-fit with a Salpeter IMF.\cite{canameras:17}

The combination of lensing with kinematics can break some of the degeneracies between
the dark matter profile and the stellar mass, as has been demonstrated at lower
redshifts.\cite{auger:10} This should work particularly well for JWST-ER1 as the effective radius of the galaxy is a factor of 3.5 smaller than the Einstein radius. 
Future NIRSpec observations of JWST-ER1 could provide
the velocity dispersion of the galaxy, as well as pin down
the redshifts of the lens and source.




\begin{methods}

\section{Discovery}

JWST-ER1 is located in the COSMOS-Web JWST data,\cite{cosmosweb} as
described in the main text.
We reduced and aligned the NIRCam images with a software pipeline that was previously developed for \textit{Hubble Space Telescope} (HST) imaging and was modified for the JWST instruments.\cite{valentino:23}
Existing HST/ACS F814W imaging
from the original COSMOS project\cite{koekemoer:07} and datasets at other wavelengths were processed in the same way, so that all space-based images are aligned to a common astrometric frame. The galaxy was found from a visual inspection
of a mosaic that was generated from the F115W, F277W, and F444W data.

It is not the first Einstein ring that was found in the COSMOS field; there are at least
two others, along with several more candidates.\cite{faure:08}
This raises the question why JWST-ER1 had not been noticed before. The main reason is that both
the source and the lens are faint in the optical, and that existing HST data in the near-IR --
while showing the lens -- are not deep enough to show the source. In Fig.\ \ref{old_data.fig}
the pre-JWST high resolution data are shown: the HST/ACS F814W from the original
COSMOS program\cite{koekemoer:07}
and a short-exposure HST/WFC3 F160W image from 3D-DASH, a wide-field survey with the
drift-and-shift (DASH) technique.\cite{mowla:22} With the benefit of hindsight the
characteristics of an Einstein ring can be glimpsed: a compact red galaxy
near the center of a blue ring.

\begin{figure}[hbtp]
  \begin{center}
  \includegraphics[width=0.95\linewidth]{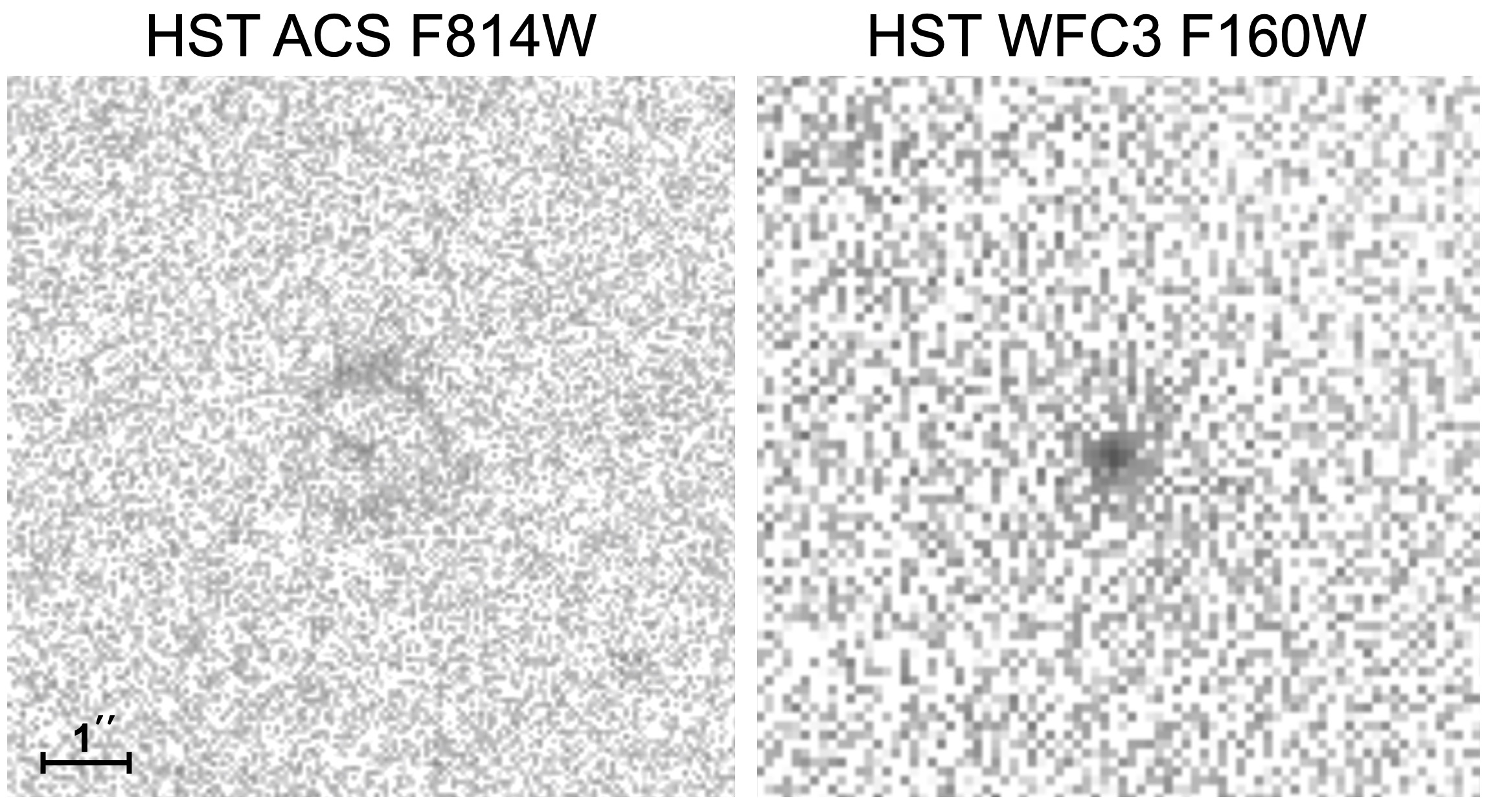}
  \end{center}
    \vspace{-0.3truecm}  
    \caption{\small \textbf{HST data of JWST-ER1.} 
Pre-JWST high resolution imaging shows the ring in ACS F814W and the galaxy in
WFC3 F160W. The system could have been flagged as a candidate Einstein ring. 
   }
   \label{old_data.fig}
    \vspace{-12pt}
\end{figure}

\section{Is it a lens?}

We consider
the possibility that the system is not a gravitational lens
but a ring galaxy, such as Hoag's object.\cite{hoag:50} Star forming
rings can be created in head-on collisions\cite{appleton:96} and there is a 
small galaxy to the southwest of the ring that could be the perturber.
The most obvious argument in favor of the lensing interpretation
is that the photometric redshift of the ring is higher than that
of the central galaxy (see main text).
However, the redshift of the ring is uncertain, and it might be possible
to fit both the lens and the ring with a model at $z\approx 2.1$. 

Here we highlight the morphology of the ring. In Fig.\ \ref{knots.fig}
we show a zoomed-in, high contrast
color image generated from the F150W and F444W data,
after subtracting the best-fitting model for the central galaxy.
There are several symmetries in the image: besides the two
bright red knots it appears that two blue knots 
are also multiple-imaged. The most compelling argument for lensing
is the morphology of the red knots (presumably the bulge of the lensed
galaxy): they are stretched into mirrored
arcs on each side of the galaxy, something that cannot be explained
in collisional ring scenarios.

\begin{figure}[hbtp]
  \begin{center}
  \includegraphics[width=0.95\linewidth]{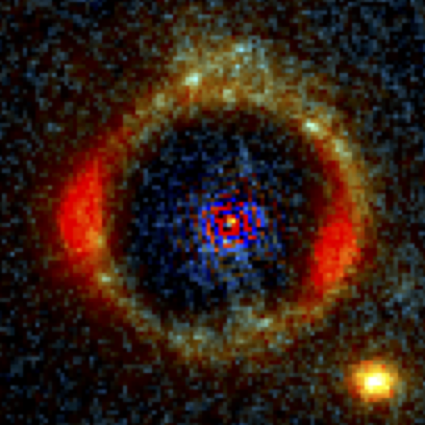}
  \end{center}
    \vspace{-0.3truecm}  
    \caption{\small \textbf{Symmetries in the ring.} 
This color image was created from the F150W and F444W images, after
subtracting the {\tt galfit} model of the central galaxy. The image spans $2.5'' \times 2.5''$.
There are several candidate multiply-imaged features along the ring.
The two red knots are very bright in F444W and are stretched into
mirrored arcs. This would be difficult to explain by any
other mechanism than gravitational lensing.
   }
   \label{knots.fig}
    \vspace{-12pt}
\end{figure}

\section{Structural parameters}

The lens galaxy is fit with the {\tt galfit} code\cite{galfit} to
determine its structure and in preparation for measuring its photometry.
We use cutouts of $4.1'' \times 4.1''$ with $0.025''$\,pix$^{-1}$ sampling
in the NIRCam bands and $0.05''$\,pix$^{-1}$ sampling in the ACS F814W
band.
The presence of the ring makes it difficult to measure the size,
Sersic index, and background level simultaneously. We therefore first measure
the background level in each band from the
outer edge of the cutout, iteratively rejecting outlying pixels, and subtract this value. Next a mask is created for the ring, by selecting pixels
in the ring area
above a flux threshold and then expanding the mask
using a $5\times 5$ pixel kernel.

The fit is performed on the F115W, F150W, F277W, and F444W images (the S/N ratio in the F814W image is too low for a stable fit). Free parameters are
the position, Sersic index, effective radius, total magnitude, axis ratio,
and position angle. We use the WebbPSF
tool\footnote{https://www.stsci.edu/jwst/science-planning/proposal-planning-toolbox/psf-simulation-tool}
to create point spread functions (PSFs) for each filter and position.
We verified that a well-exposed nearby star does not lead to qualitatively
different results.

The structural parameters are listed in Table 1.
The parameters in the four bands are in good agreement, despite
the factor of four range in wavelength and resolution going from F115W
to F444W.  The average effective radius $r_{\rm e} = 8.8\pm 0.8$ pixels,
or $0.22''\pm 0.02''$, where the rms of the four individual measurements
is taken as the uncertainty. The Sersic index $n=5.0\pm 0.6$.
The axis ratio is very close to 1 and there is no
consistent position angle between the bands; in what follows
we therefore assume that $b/a=1.0$.

\begin{small}
\begin{table}
\centering
 \begin{tabular}{c c c c c} 
 \hline
 Filter & $r_{\rm e}$ [pix] & $n$ & $b/a$ & PA \\ 
 \hline\hline
${\rm F115W}$ & $7.9\pm 0.7$ & $4.1\pm 0.3$ & $0.94$ & $77$\\
${\rm F150W}$ & $9.9\pm 0.5$ & $4.9\pm 0.2$ & $0.96$ & $-15$\\
${\rm F277W}$ & $8.8\pm 0.4$ & $5.3\pm 0.2$ & $0.98$ & $-16$\\
${\rm F444W}$ & $8.6 \pm 0.4$ & $5.6\pm 0.2$ & $0.99$ & $-23$\\
 \hline
 \end{tabular}
\caption{Structural parameters of the lens.}
\end{table}
\end{small}

\section{Photometry}

Total magnitudes of the lens are determined by fitting 
the five bands (now including ACS F814W) with
{\tt galfit}, holding all parameters except the total magnitude fixed
to the average values determined above. This constrained (or forced)
fit ensures that the relative fluxes between the bands
are measured in a self-consistent way, and
not compromised by PSF or aperture effects. The results are listed
in Table 2, with 0.05\,mag systematic error added in quadrature to the
random errors.
For the comparison of the lensing mass to the stellar
mass it is not the total flux but the projected flux within the
Einstein radius that matters. Using a model profile that is not
convolved with the PSF we determine that $79$\,\% of the
total flux is within the Einstein radius. For convenience
the magnitudes within
the Einstein radius are listed in a separate column in Table 2.
We tested that simple aperture photometry on the galaxy, with the ring masked, gives a redshift and $M/L$ ratio that are within the uncertainties of the fiducial values.

\begin{small}
\begin{table}
\centering
 \begin{tabular}{c c c} 
 \hline
 Filter & Total &  In ring\\
 \hline\hline
${\rm F814W}$ & $25.97\pm 0.14$ & $26.22\pm 0.14$\\
${\rm F115W}$ & $23.36\pm 0.08$ & $23.61\pm 0.08$\\
${\rm F150W}$ & $22.14\pm 0.05$ & $22.39\pm 0.05$\\
${\rm F277W}$ & $20.95\pm 0.05$ & $21.20\pm 0.05$\\
${\rm F444W}$ & $20.43\pm 0.05$ & $20.68\pm 0.05$\\
 \hline
 \end{tabular}
\caption{Photometry of the lens (AB mag).}
\end{table}
\end{small}

Photometry of the ring is performed by simply summing the
flux in apertures.
Two apertures are used: one covering both of the red concentrations
within the ring, and one covering the rest of the ring. No attempt
is made to correct for the PSF variation between bands, but
the apertures are purposefully made large enough to mitigate these effects.
We use the photometry of the ring to derive an approximate redshift, and we caution against using it to determine
detailed stellar population parameters of the lensed galaxy.
The magnitudes for the two apertures are listed in Table 3.

\begin{small}
\begin{table}
\centering
 \begin{tabular}{c c c} 
 \hline
 Filter & Blue ring &  Red knots\\
 \hline\hline
${\rm F814W}$ & $24.55\pm 0.06$ & $27.53\pm 0.35$\\
${\rm F115W}$ & $23.85\pm 0.09$ & $26.66\pm 0.29$\\
${\rm F150W}$ & $23.24\pm 0.07$ & $25.40\pm 0.11$\\
${\rm F277W}$ & $22.31\pm 0.05$ & $23.58\pm 0.05$\\
${\rm F444W}$ & $21.89\pm 0.05$ & $22.70\pm 0.05$\\
 \hline
 \end{tabular}
\caption{Photometry of the source (AB mag).}
\end{table}
\end{small}

\section{Prospector fits}

The redshift of the lens and its stellar population parameters are determined
jointly using the \texttt{Prospector} inference framework;\cite{johnson:21} specifically the Prospector-$\alpha$ model\cite{leja:17} and the MIST stellar isochrones\cite{choi:16,dotter:16} from FSPS.\cite{conroy:10}
Prospector-$\alpha$ describes the star formation history (SFH) non-parametrically via mass formed in seven logarithmically-spaced time bins, and assumes a continuity prior to ensure smooth transitions between bins.\cite{leja:19nonpar}
We additionally adopt a dynamic SFH($M,z$) prior\cite{wang:23} which follows the observed cosmic star formation rate density, favoring rising SFHs in the early universe and falling SFHs in the late universe, with a mass-based adjustment to reflect downsizing.
The model consists of 18 free parameters, including the form of the attenuation curve, and sampling is performed using the dynamic nested sampler \texttt{dynesty}.\cite{speagle:20} 
The parameters for the lens are determined from the photometry inside the ring.
We report the posterior median of the inferred physical parameters in Table 4, assuming a Chabrier IMF. The uncertainties reflect the 16$^{\rm th}$ and 84$^{\rm th}$ percentiles.

The uncertainties in the redshift and mass may
seem suspiciously 
small given that we only five photometric
datapoints. The reason why the key parameters are so well constrained is that the photometry tells us only one thing, but it does so precisely:
there is a large break in the
spectral energy distribution at $1.2\,\mu$m. 
The constraints on the redshift and $M/L$ ratio
follow directly from this. We performed two robustness
tests to determine how sensitive the results are to the specifics of our methodology. First, 
removing the SFH prior leads to negligible differences to the redshift and mass, and all the posterior medians are consistent within $1\sigma$. The only notable change is that the prior decreases the uncertainty on the star formation rate. This behavior is expected: at this redshift and mass the prior prefers a  falling SFH, consistent with the observed high mass (i.e., high previous star formation rate) and low current star formation rate. Second, determining the
redshift with the {\tt eazy} code\cite{eazy} (which uses a pre-rendered set of templates) gives
$z_{\rm phot}=1.91^{+0.18}_{-0.17}$ and no viable secondary solutions, in good
agreement with our fiducial value.

The lensed galaxy is modelled in the same way as the lens, except that the scale-dependent SFH prior is not included due to the lensing magnification.
The main goal is to determine the redshift of the lensed galaxy.
For completeness we list stellar population parameters
for the two apertures on the ring as well in Table 4, although they are not used in the analysis.

\begin{small}
\begin{table}
\centering
 \begin{tabular}{l c c c} 
 \hline
 & Lens & Blue ring & Red knots  \\
 \hline\hline
$z$ & $1.94^{+0.13}_{-0.17}$ & $2.89^{+0.27}_{-0.98}$ & $2.98^{+0.42}_{-0.47}$ \\
{\rm log M/M$_{\odot}$} & $11.03^{+0.09}_{-0.13}$ & $10.43^{+0.22}_{-0.56}$ & $10.63^{+0.16}_{-0.18}$ \\
{\rm SFR [}\msun{}\,{\rm yr}$^{-1}${\rm ]} & $4^{+19}_{-3}$ & $64^{+36}_{-25}$ & $19^{+48}_{-16}$ \\
{\rm age [Gyr]} & $1.9^{+0.3}_{-0.6}$ & $0.8^{+0.3}_{-0.5}$ & $1.1^{+0.4}_{-0.4}$ \\
 \hline
 \end{tabular}
\caption{Inferred parameters.}
\end{table}
\end{small}

\section{Obscured star formation}

The low star formation rate of JWST-ER1g derived above
implies a low gas surface density, and
hence a low contribution of gas to the total mass budget within the
Einstein ring. However, the Prospector fits do not provide strong constraints
on the amount of star formation that is optically-thick.
The field has been observed with Spitzer/MIPS, as part of the S-COSMOS survey,\cite{sanders:07}  and we use the 24\,$\mu$m data to assess whether
JWST-ER1g has a hidden obscured star burst.

The S-COSMOS 24\,$\mu$m image is shown in Fig.\ \ref{mips.fig}. The
galaxy is not detected. We determine an upper limit to the star formation
rate from a redshift-dependent relation between observed 24\,$\mu$m flux
and total IR luminosity that was calibrated with Herschel
data.\cite{wuyts:11,whitaker:12} The $3\sigma$ upper limit is
63\,\msun\,yr$^{-1}$.

\begin{figure}[hbtp]
  \begin{center}
  \includegraphics[width=0.95\linewidth]{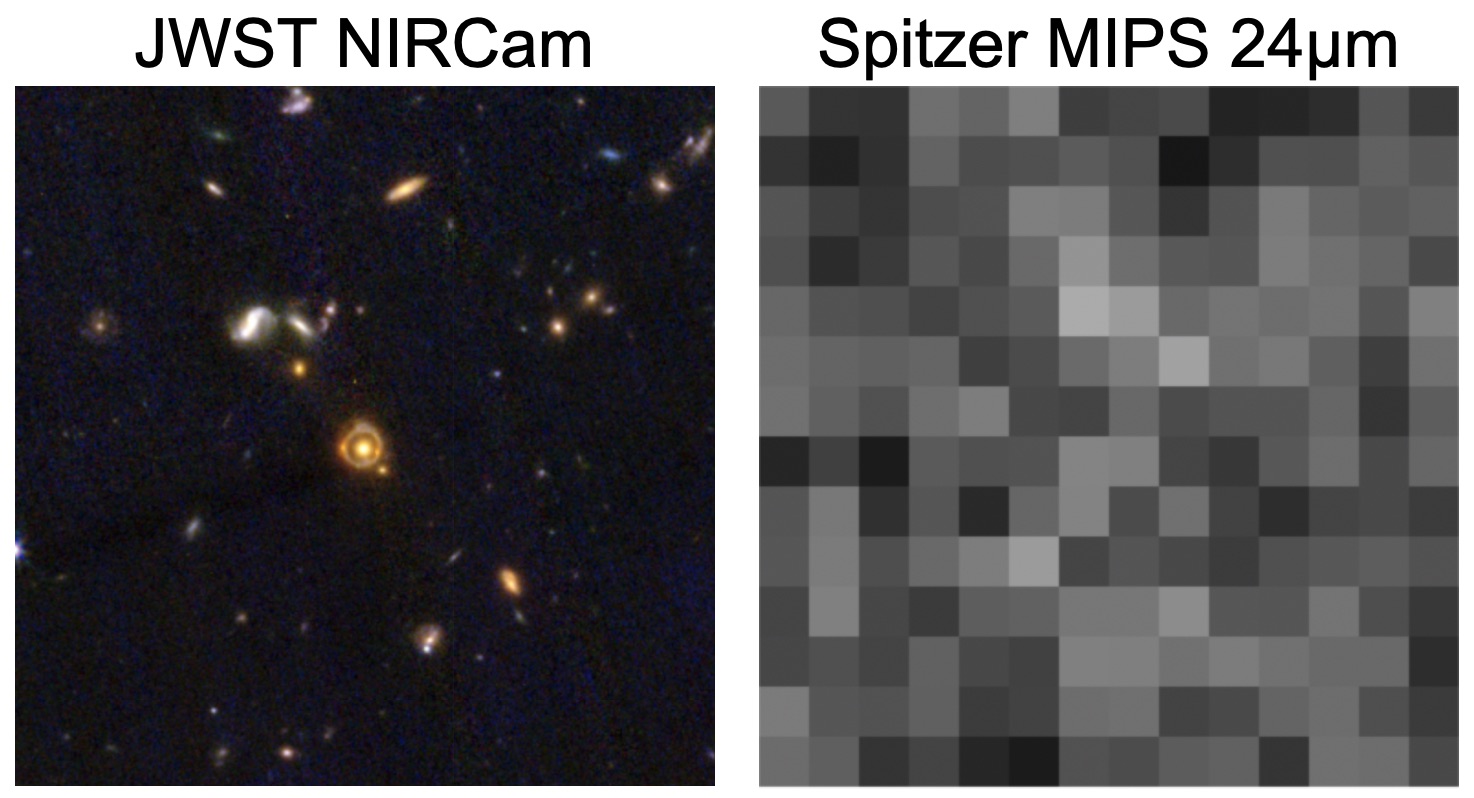}
  \end{center}
    \vspace{-0.3truecm}  
    \caption{\small \textbf{Non-detection at 24 micron.} 
The $30'' \times 30''$ region around JWST-ER1 as observed with
Spitzer/MIPS at 24\,$\mu$m. The galaxy is not detected, and the
upper limit on the star formation rate of the lensing galaxy
is 63\,\msun\,yr$^{-1}$.
   }
   \label{mips.fig}
    \vspace{-12pt}
\end{figure}

\section{Comparison to other $z\approx 2$ galaxies}

As noted in the main text, JWST-ER1g is a typical example of the class
of massive, quiescent $z\approx 2$ galaxies. This is demonstrated explicitly
in Fig.\ \ref{relations.fig}. Fig.\ \ref{relations.fig}a shows
that the galaxy falls in the quiescent region of the UVJ diagram.
The boundaries are the averages of the $z=1.75$ and $z=2.25$ limits
determined for the NEWFIRM Medium Band Survey.\cite{whitaker:11}
It is relatively red within the quiescent region,
indicating an old age and/or some dust, as also implied by the Prospector
fit. In Fig.\ \ref{relations.fig}b the galaxy's size is compared to
the canonical size-mass relations\cite{wel:14} for  quiescent and
star forming galaxies, again taking the average
of the listed relations for $z=1.75$ and $z=2.25$. The galaxy falls
on the relation for quiescent galaxies.

\begin{figure}[hbtp]
  \begin{center}
  \includegraphics[width=1.0\linewidth]{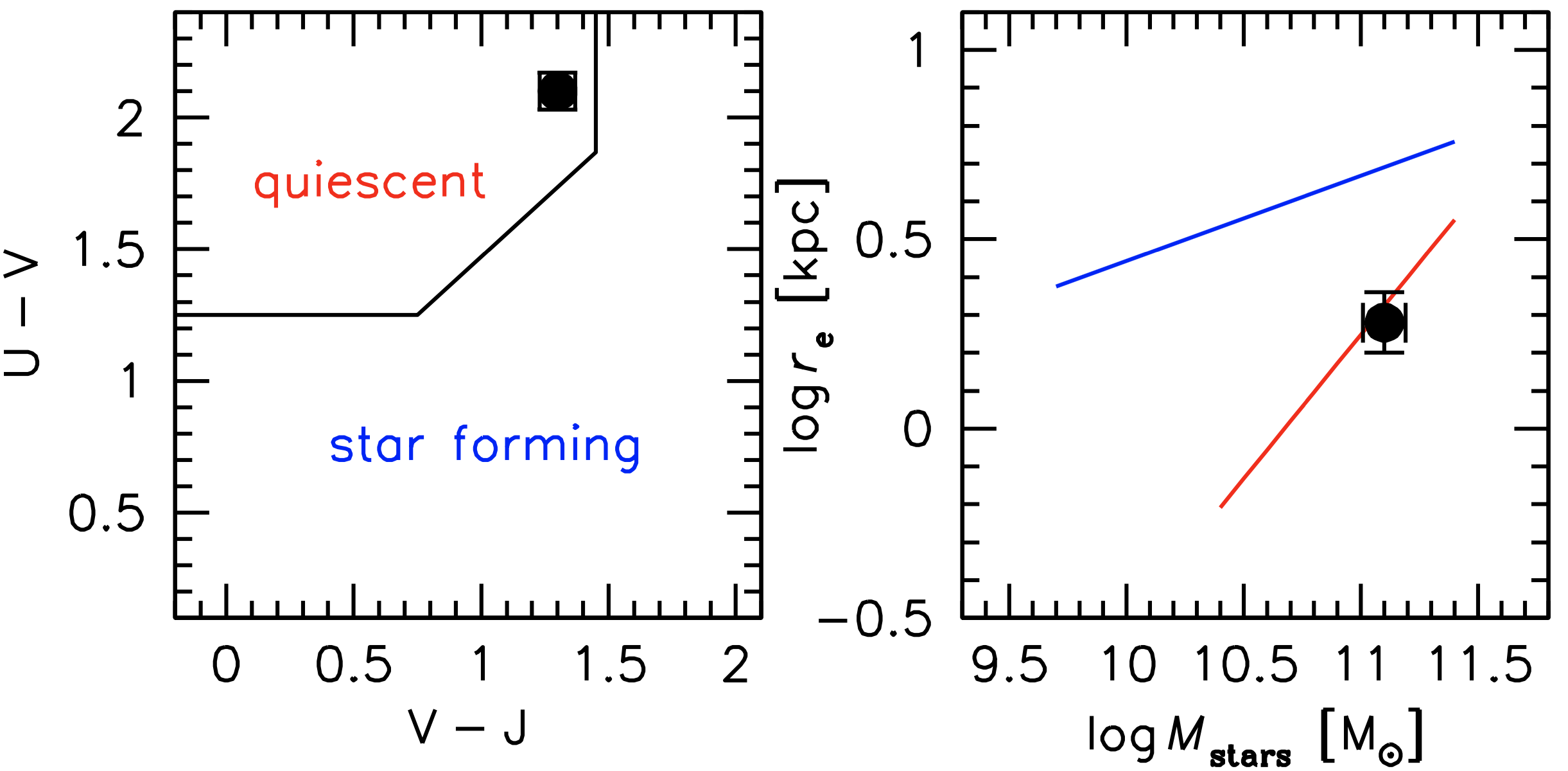}
  \end{center}
    \vspace{-0.3truecm}  
    \caption{\small \textbf{JWST-ER1g in context.} 
a) Location in the UVJ diagnostic diagram for $z\sim 2$. The galaxy falls in the
quiescent region.\cite{whitaker:11} b) Location in the size-mass diagram for $z\sim 2$.
It falls on the size-mass relation for quiescent galaxies.\cite{wel:14}
Data points are measurements $\pm$\,sd.
   }
   \label{relations.fig}
    \vspace{-12pt}
\end{figure}

\section{Lensing mass}

The mass within the Einstein radius is given by
\begin{equation}
M(<\theta) = \frac{\theta^2 c^2 D_{\rm l} D_{\rm s}}{4G D_{\rm ls}},
\label{lens.eq}
\end{equation}
with $\theta$ the observed Einstein radius in radians, $D_{\rm l}$ the angular
diameter distance to the lens,  and $D_{\rm s}$ the angular diameter distance to
the source. The  parameter $D_{\rm ls}$
is the distance between the lens and the source,
which is
\begin{equation}
D_{\rm ls} = D_{\rm s} - \frac{1+z_{\rm l}}{1+z_{\rm s}} D_{\rm l}
\end{equation}
in a flat Universe.\cite{hogg:99}
The uncertainties are determined numerically, by drawing values of
$z_{\rm s}$, $z_{\rm l}$, and $\theta$ from their probability distributions and
calculating $M(<\theta)$ for each set of draws. 

The high lens mass is driven by the large diameter of the Einstein
ring combined with the relatively high redshift of the lens. 
Forcing $z_{\rm l}=1.5$ (which is outside of the full posterior
distribution of 5000 samples) lowers the mass to
$M_{\rm lens}=4.1\times 10^{11}$\,\msun, but also
lowers the derived Chabrier stellar mass
to $M_{\rm stars}=
0.6\times 10^{11}$\,\msun. The ratio of
lensing mass to Chabrier mass is $\approx 7$,
very similar to the results for $z=1.94$.

The source redshift is the most uncertain parameter in Eq.\ \ref{lens.eq}, but the lensing mass is not very sensitive to it. The lensing mass is lower for higher source redshifts, but is $3.4\times 10^{11}$\,\msun\ even for $z_{\rm s}=10$. 
The main effect of the uncertainty in the source redshift is that it causes
an asymmetry in the error distribution of $M_{\rm lens}$, with a tail to very high
masses. This is because
the mass increases rapidly when $z_{\rm s} \approx z_{\rm l}$: the mass is
$>10^{12}$\,\msun\ if $z_{\rm s}<2.5$, and reaches $4\times 10^{12}$\,\msun\
for $z_{\rm s}=2.1$.

\section{Dark matter contribution}

The projected dark matter mass within the ring can be calculated
by integrating an NFW profile\cite{navarro:97} along a cylinder with
a radius of 6.6\,kpc.\cite{lokas:01} The scaling $\log c = 0.81 - 0.09(\log M_{\rm vir} -12)$ is used to determine the concentration as a function of halo
mass.\cite{dutton:14} The resulting relation between projected dark
matter mass
within the ring and total halo mass is shown in Fig.\ \ref{halomass.fig}.

The relation is shallow, due to the decreasing concentration with halo mass. We estimate the dark matter contribution to the
lensing mass from the
halo mass -- stellar mass relation.\cite{behroozi:13b} 
We find $M_{200} = 1.0^{+2.6}_{-0.5} \times 10^{13}$\,\msun, with
the relatively large uncertainty driven by the steepness of the relation
in this regime. The corresponding projected dark matter mass within $6.6$\,kpc
is  $M_{\rm dm} = 2.6^{+1.6}_{-0.7}
\times 10^{11}$\,\msun\ for
an NFW halo. 

\begin{figure}[hbtp]
  \begin{center}
  \includegraphics[width=0.95\linewidth]{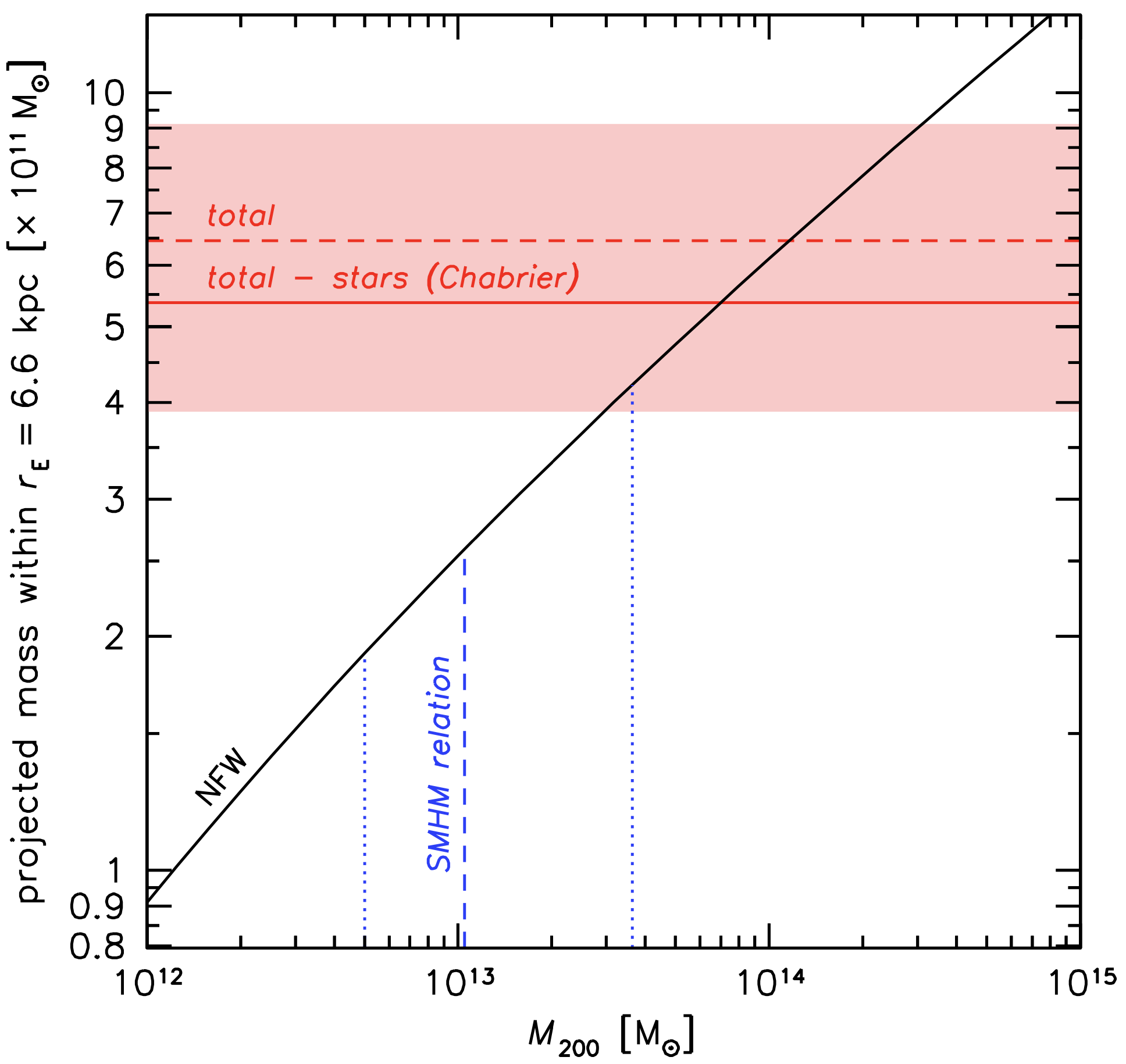}
  \end{center}
    \vspace{-0.3truecm}  
    \caption{\small \textbf{Relation between projected mass within
    the Einstein radius and total halo mass.} 
The dashed horizontal line indicates the total lensing mass. The solid horizontal line is the remaining mass after subtracting the stellar mass, for a Chabrier IMF and with the band indicating the $\pm 1\sigma$ uncertainty. 
The blue vertical lines show the expected
halo mass from the $z=2$ stellar mass -- halo mass (SMHM) relation\cite{behroozi:13b} and its uncertainty.
The solid black line is the expected relation
for NFW halos at $z=1.94$. The galaxy has more mass
within the Einstein radius than expected
from a Chabrier IMF, the SMHM relation, and
an NFW profile.
   }
   \label{halomass.fig}
    \vspace{-0pt}
\end{figure}

\begin{figure}[hbtp]
  \begin{center}
  \includegraphics[width=0.95\linewidth]{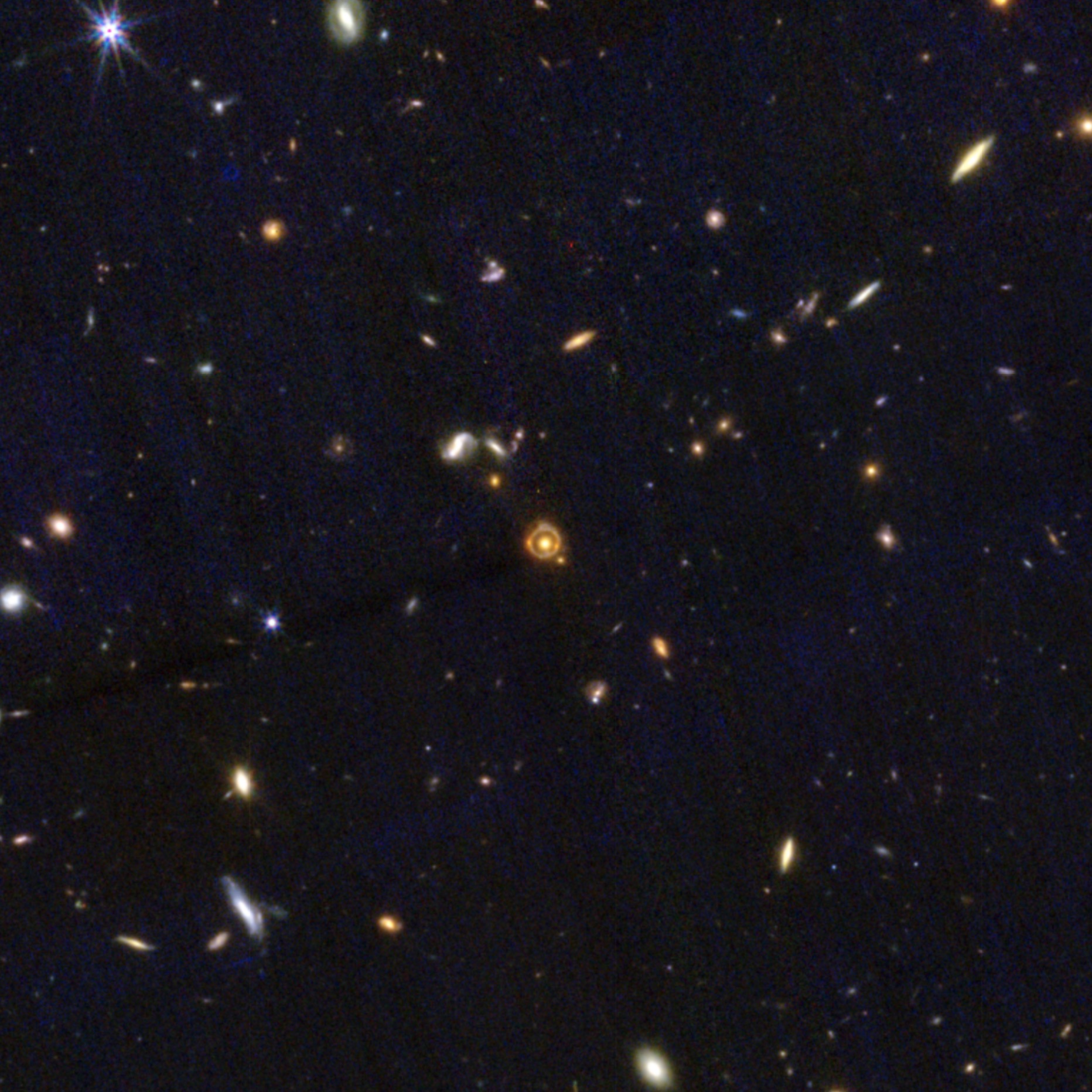}
  \end{center}
    \vspace{-0.3truecm}  
    \caption{\small \textbf{Environment of the lens.} 
The $1' \times 1'$ region of the COSMOS-Web mosaic centered on JWST-ER1.
The object appears to be relatively isolated; it may of course
be part of a group, but
it is not near the central regions of a rich cluster.
   }
   \label{environ.fig}
    \vspace{-12pt}
\end{figure}

The solid horizontal line indicates the difference between the lensing mass and
the stellar mass of JWST-ER1g, for a Chabrier IMF. 
To explain the missing
mass entirely with dark matter the NFW halo mass would have to be $\approx 7\times 10^{13}$\,\msun. Halos of this mass at $z=2$ are progenitors of clusters
at $z=0$.
The number density of
halos with $M_{200}>7\times 10^{13}$\,\msun\ at $z=1.94$
is $2\times 10^{-7}h^{-3}$\,Mpc$^{-3}$, corresponding
to 1.4 in the redshift range $1.75<z<2.25$ in the 0.35\,deg$^2$ of
the available COSMOS-Web area.\cite{murray:21}
Halos
with slightly lower masses are of course more common, and still consistent
with the lensing constraints. The lower $1\sigma$ bound
on the lensing mass corresponds to a halo mass of
$M_{200}>3\times 10^{13}$\,\msun\ (see Fig.\ \ref{halomass.fig}), 
and there are $\sim 15$ such halos in the COSMOS-Web area.

\section{Environment of JWST-ER1}

Gravitational lensing is sensitive to the weighted integral of all mass between the
source and the observer, and we briefly consider whether nearby galaxies or structures along the line of sight
could contribute to the mass. We also consider whether
JWST-ER1g is the central galaxy of the progenitor
of a cluster (see above).
The immediate environment of JWST-ER1 is shown in Fig.\ \ref{environ.fig},
generated from the NIRCam F115W, F277W, and F444W images. The region does not stand out in any way; the galaxy is either isolated or in a sparse group, but not in a massive cluster. Furthermore, there are no other bright galaxies projected along
the line of sight. We infer that the contributions from
other galaxies to the
$6.7\times 10^{11}$\,\msun\ mass within the Einstein radius
are almost certainly negligible.

\end{methods}

\vspace{3pt}
\noindent\rule{\linewidth}{0.4pt}
\vspace{3pt}





\begin{addendum}
\item[Data availability]
The COSMOS-Web data are publicly available from the STScI MAST Archive.
\item[Code availability]
We have made use of standard data analysis software in the Python
and IRAF environments, and the publicly available code
{\tt galfit}.\cite{galfit}
 \item[Acknowledgements]
This project is based on data from the JWST Cycle 1
COSMOS-Web Treasury program.
Support from STScI grants GO-16259 and GO-16443 is gratefully acknowledged. We are grateful to the referees who provided outstanding feedback and corrected an important error in the original manuscript.
 \item[Author Contributions] P.v.D.\ and G.B.\ identified the
 galaxy.  P.v.D.\ led the analysis and wrote the
 manuscript. G.B.\ reduced the data and produced the mosaic.
 B.W.\ and J.L.\ performed the Prospector analysis.  All authors aided in the analysis and interpretation and contributed
 to the final manuscript.
 \item[Competing Interests] The authors declare that they have no
   competing financial interests. Correspondence and requests for
   materials should be addressed to P.v.D.~(email:
   pieter.vandokkum@yale.edu).

\end{addendum}

\vspace{3pt}
\noindent\rule{\linewidth}{0.4pt}
\vspace{3pt}

\bibliography{master}



\appendix




\end{document}